\documentclass[sigconf]{acmart}
\usepackage{xcolor}
\usepackage{colortbl}

\AtBeginDocument{%
  }

\usepackage{framed}
\usepackage{tcolorbox}
\definecolor{awesome}{RGB}{70,130,180}
\usepackage{graphicx}
\usepackage{enumitem} 
\usepackage{float} 

\setcopyright{acmlicensed}
\copyrightyear{2026}
\acmYear{2026}
\acmDOI{XXXXXXX.XXXXXXX}

\acmConference[EASE 2026]{The 30th International Conference on Evaluation and Assessment in Software Engineering}{Tue 9 - Fri 12 June 2026}{Glasgow, United Kingdom}

\acmISBN{978-1-4503-XXXX-X/2018/06}

\newcommand{\RqOne}{\textbf{RQ1:} \textbf{In the development process, who references the migration guide, where in the pull request do they reference it, and at what level of detail?}}

\newcommand{\RqTwo}{\textbf{RQ2:} \textbf{For what purposes and in what contexts do developers reference the migration guide?}}

\definecolor{awesome}{RGB}{70,130,180}
\makeatletter
\newsavebox{\acmbox@box}
\newlength{\acmbox@innerwd}
\newenvironment{acmbox}[1]{%
  \def\acmbox@title{#1}%
  \par\medskip\noindent
  \setlength{\fboxsep}{6pt}%
  \setlength{\fboxrule}{0.8pt}%
  \setlength{\acmbox@innerwd}{\dimexpr\linewidth-2\fboxsep-2\fboxrule\relax}%
  \begin{lrbox}{\acmbox@box}%
  \begin{minipage}{\acmbox@innerwd}%
  \setlength{\parindent}{0pt}%
  \setlength{\parskip}{0.35\baselineskip}%
  {\color{awesome}\bfseries \acmbox@title}\par\vspace{2pt}%
}{%
  \end{minipage}%
  \end{lrbox}%
  \noindent\fcolorbox{awesome}{gray!5}{\usebox{\acmbox@box}}%
  \par\medskip
}
\makeatother

\begin{document}

\title{How Do Developers Use Migration Guides? \\A Case Study of Log4j}
\renewcommand{\shorttitle}{How Do Developers Use Migration Guides? A Case Study of Log4j}

\author{Takahiro Monno}
\orcid{0009-0001-4669-8411}
\affiliation{%
  \institution{Nara Institute of Science and Technology}
  \city{Ikoma}
  \state{Nara}
  \country{Japan}
}
\email{monno.takahiro.mv3@naist.ac.jp}

\author{Kazumasa Shimari}
\orcid{0000-0001-8837-5090}
\affiliation{%
  \institution{Wakayama University}
  \city{Wakayama}
  \state{Wakayama}
  \country{Japan}
}
\email{shimari@wakayama-u.ac.jp}

\author{Tetsuya Kanda}
\orcid{0000-0003-4620-7050}
\affiliation{%
  \institution{Notre Dame Seishin University}
  \city{Okayama}
  \state{Okayama}
  \country{Japan}
}
\email{kanda@m.ndsu.ac.jp}

\author{Kazuma Yamasaki}
\orcid{0009-0001-2657-8682}
\affiliation{%
  \institution{Nara Institute of Science and Technology}
  \city{Ikoma}
  \state{Nara}
  \country{Japan}
}
\email{yamasaki.kazuma.yj9@naist.ac.jp}

\author{Kenichi Matsumoto}
\orcid{0000-0002-7418-9323}
\affiliation{%
  \institution{Nara Institute of Science and Technology}
  \city{Ikoma}
  \state{Nara}
  \country{Japan}
}
\email{matumoto@is.naist.jp}

\begin{abstract}
Migration guides are a form of software documentation that helps developers address breaking changes introduced in library version updates. Prior studies have examined documents such as release notes, API reference manuals, and patch notes. However, research that focuses specifically on migration guides remains limited. Improving the usability and coverage of migration guides is essential for helping developers resolve breaking changes efficiently. Yet, we still lack a clear understanding of how migration guides are currently provided and how developers use them in practice.

To fill this gap, we first investigate whether libraries known to introduce incompatibilities provide migration guides. We then conduct a detailed case study on Log4j, a library that has experienced large-scale breaking updates in the past. We empirically analyze how developers refer to and use the official migration guide in real-world projects.
We find that pull request authors most frequently reference the migration guide in the pull request description, and that most references (82.81\%) link to the entire guide rather than specific sections. We also find that developers use migration guides not only during major version updates but also during subsequent maintenance tasks, suggesting that the guides serve as a resource throughout the entire migration process.

\end{abstract}



\maketitle

\section{Introduction}

\label{sec:introduction}

When library providers release new versions, they sometimes introduce backward-incompatible changes, commonly called ``Breaking Changes''~\cite{Jayasuriya2024}.
Such breaking changes can render client systems defective or alter their behavior, so developers need to address them whenever they update library versions~\cite{venturini2023}.
Migration guides are documents that summarize the changes between two versions of a library and provide concrete instructions and replacement mappings to help developers update their code~\cite{Li2013}.
By following migration guides, developers can obtain code-level mappings for resolving breaking changes in their own projects~\cite{Ko2014, Kao2022}.

Prior empirical studies have shown that developers often rely on software documentation and external web resources to understand and recover from breaking changes.
For example, Zampetti et al. found that documentation URLs were the second most frequently referenced category of external URLs in GitHub PRs, indicating that developers often share documentation links to support team members during change tasks~\cite{Zampetti2017}.
Among such documentation, migration guides play an important role because they directly connect old or changed APIs to their new versions and provide clear steps to adapt the code~\cite{Li2013}.
Despite this practical importance, we still lack a clear understanding of how migration guides are currently provided and how developers use them in practice.

Existing studies on documentation that supports library migration have focused on the provider's perspective, analyzing the content and availability of such documentation.
For example, Li et al. reported that migration guides for web service APIs do not always cover all changes introduced in a version update and may even contain omissions and errors~\cite{Li2013}.
Yasmin et al. examined deprecation practices of RESTful APIs and found that not all providers supply replacement messages directing developers to alternative endpoints~\cite{Yasmin2020ICSME}.
While these findings clarify what providers offer to support migration, no empirical study has examined how client developers use migration guides within version update workflows.

In this paper, we conduct the first empirical study investigating how developers use migration guides when coping with breaking changes in library dependencies.
First, we examine to what extent library providers publish migration guides compared with release notes, thereby quantifying the availability of migration guides.
Second, we conduct a case study of Log4j, a widely used logging library that introduced extensive breaking changes between version 1 and version 2, to analyze how developers reference the official migration guide in practice when updating their client software.
Together, these empirical analyses provide initial evidence on how migration guides are provided and used in practice. We formulate the following research questions:

\noindent
\RqOne\\
We empirically analyze who refers to migration guides, where the references appear, and the granularity of such references. Unlike prior work that investigates migration guides from the provider side~\cite{Yasmin2020ICSME}, we investigate migration guide usage from the client developers' perspective.

\noindent
\RqTwo\\
Multiple authors conduct a manual analysis of the PRs examined in RQ1 to deepen our understanding of how migration guides are used. Specifically, we analyze development and maintenance activities in the PRs as well as the status of library updates, and we exploratorily examine developers' purposes for referring to migration guides.

\noindent
\textbf{Contributions:}
\begin{itemize}[leftmargin=*]
    \item To the best of our knowledge, this study is the first empirical study investigating how migration guides are utilized in OSS.
    \item We find that PR authors most frequently reference the migration guide in the pull request description, and that most references (82.81\%) link to the entire guide rather than specific sections.
    \item We also find that developers use migration guides not only during major version updates but also during subsequent maintenance tasks, suggesting that the guides serve as a resource throughout the entire migration process.
\end{itemize}

\noindent
\textbf{Replication Package:} To facilitate replication and further studies, we provide the data and source code used in our study as a replication package on Zenodo\footnote{\url{https://doi.org/10.5281/zenodo.18845533}}.

\section{Related Work}
\label{sec:motivating_example}
\subsection{Importance of Migration Guides}

Migration guides are indispensable for coping with breaking changes, a routine yet burdensome part of library evolution.
In Java libraries, breaking changes are not exceptional: about 30\% of API changes introduce backward incompatibility~\cite{Xavier2017}.
Moreover, dependency updates often lag~\cite{Miller2023}; 32\% of developers avoid updating due to compatibility concerns~\cite{Derr2017}, and many projects took several months to adopt patched Log4j versions after critical vulnerabilities were disclosed~\cite{Tanaka2025}.
This lag makes breaking changes particularly costly to handle, and prior research~\cite{Li2013} shows that, by following migration guides, developers can obtain code-level mappings that help them resolve breaking changes effectively.
Accordingly, migration guides play a crucial role in supporting developers during version updates.
Therefore, migration guides provide an important means to support developers when they face breaking changes.

\subsection{Existing Studies on Migration Guides}

Existing work has examined migration guides and software documentation from several angles.
Li et al. reported that official migration guides for web service APIs do not always cover all changes introduced in a version update and may even contain omissions and errors, indicating that migration guides are not always complete or accurate~\cite{Li2013}.
For example, in Sina Weibo API, they found an omission and a mistake in the migration guide.
Ko et al. analyzed 260 API documents for deprecated Java APIs and found that 39\% of them did not provide information about alternative APIs, suggesting that important migration-related elements are often missing and may hinder users' update tasks~\cite{Ko2014}.
Brito et al. further showed that, when coping with difficult breaking changes, library users frequently turn to community Q\&A sites to search for concrete solutions, especially when official documentation is insufficient~\cite{Brito2020}. 

These existing works collectively suggest that official documentation often fails to provide sufficient guidance for migration tasks. 
However, they either analyze software documentation broadly without isolating migration-specific guides or focus on the provision of information from the authors' side. 
Our study focuses exclusively on migration guides, which play a crucial role in addressing breaking changes.
From the perspective of developers, our study aims to identify key issues in migration guides by analyzing how they are utilized within actual development workflows.

\section{Preliminary Study}
\label{sec:pre}
\subsection{Purpose}

This preliminary study investigates the current landscape of documentation supporting library version updates. 
Prior studies~\cite{Yasmin2020ICSME} have reported the limited availability of migration guides for deprecated APIs. In preparation for our empirical investigation into migration guide usage, we conducted a preliminary survey to verify the scarcity of such guides.
In this study, we focused on migration guides and release notes as the primary sources of documentation containing information regarding library updates.

\subsection{Method}

We analyzed a dataset of breaking change instances that prior research collected~\cite{Reyes2024}. The dataset includes 571 instances of breaking changes from 153 Java projects.

We conducted the analysis as follows.
First, we extracted all libraries that introduced breaking changes from the dataset.
We then examined the GitHub repositories and official websites of each library to confirm whether they provide migration guides or release notes.
In this study, we define migration guides as documentation that provides version update information separately from release notes.
For libraries that belong to the same framework, such as ``\texttt{spring-core}'' and ``\texttt{spring-context}'', we treated them as a single library group in this study.

\subsection{Result and Implications}

Table~\ref{tab:Provision ratio} shows the number of libraries that provide migration guides.
Among 101 distinct libraries, 28 provide both migration guides and release notes. In contrast, 65 provide release notes but do not provide migration guides. 8 provide neither migration guides nor release notes. No library provides guides without release notes.

\begin{table}[tb]
\centering
\caption{Number of Libraries Providing Migration Guides and Release Notes}
\label{tab:Provision ratio}
\begin{tabular}{l|cc}
\hline
 & \multicolumn{2}{c}{\textbf{Release Notes}} \\
\textbf{Migration Guide} & Provided & Not Provided \\
\hline
\quad Provided & 28 & 0 \\
\quad Not Provided & 65 & 8 \\
\hline
\end{tabular}
\end{table}

These results show that 92.08\% of libraries that introduce breaking changes provide release notes. However, only 27.72\% provide migration guides.
This finding suggests that libraries provide migration guides less frequently than release notes even when they introduce breaking changes, indicating that the current provision of migration guides remains insufficient. The environment that supports library users in handling breaking changes has not yet reached a sufficient level.

\section{Study Design}
\label{sec:proposed_method}
The objective of this study is to clarify how developers use migration guides in real-world projects.
In the preliminary survey described in the previous section, we found that many libraries that introduce breaking changes do not provide migration guides.
However, for libraries that do provide migration guides, we still do not understand how developers actually use them in practice.

In this section, we design a case study to analyze how developers use a migration guide in practice.

We focus on Log4j, which provides a migration guide and is widely known for large breaking changes between version 1 and version 2. We conduct an empirical analysis of how developers refer to and use its migration guide.

\subsection{Five Perspectives on Migration Guide Usage}

We analyze migration guide usage from the following five perspectives in order to understand actual usage patterns.

\begin{enumerate}
\item \textbf{Actor of reference} (PR author vs. reviewer)
\item \textbf{Location of reference} (PR description vs. comment)
\item \textbf{Level of reference} (entire guide vs. specific section)
\item \textbf{Type of development and maintenance activity}
\item \textbf{Library update status} (major update vs. no update)
\end{enumerate}

Regarding (1) and (2), prior work emphasizes that we need to examine the role of developers and the context in which they work when they require information from external resources~\cite{Ponzanelli2014}.
Regarding (3), we examine whether developers use fragment identifiers when they share links to the migration guide. A fragment identifier moves the reader directly to a specific section of a web page. Prior research reports that programmers often avoid careful reading of documents because they struggle to locate the needed information~\cite{Novick2006}.
For (4), we introduce the type of development and maintenance activity as an analytical dimension to organize migration guide usage. Research on software maintenance treats the classification and motivation of change activities as important units of analysis~\cite{koyyada2022automatedopensourceassessment}. 
By distinguishing the types of changes, we aim to clarify the differences in practical support needs.

These perspectives assume that developers use migration guides for different purposes depending on the situation. Understanding these patterns is essential for designing guides that support diverse usage scenarios. Based on these perspectives, we conduct an exploratory investigation of migration guide usage.

\subsection{Data Collection}

We use the following two URLs as links to the Log4j migration guide in our study: the migration guide for the major upgrade~\footnote{\url{https://logging.apache.org/log4j/2.x/migrate-from-log4j1.html}} and its redirected page~\footnote{\url{https://logging.apache.org/log4j/2.x/manual/migration.html}}.

We collect data by using the GitHub Search API to retrieve PRs that reference the migration guide URLs. We then manually filter the collected PRs and retain only those that appropriately reference the URLs.
We exclude issues from our analysis and focus only on PRs. Some issues may also be related to library migration, but they do not always result in a certain implementation because issues are often used for raising problems or discussing future plans. In contrast, PRs contain code changes, and we can easily see how the discussion in the PRs is related to the code base. At this initial stage of our investigation into migration guide usage, we consider PRs to be an appropriate unit of analysis.

\subsection{RQ1: Quantitative Analysis of Reference Patterns}

To investigate how developers use the migration guide in PRs, we define the following dimensions.

\noindent\textbf{(1) Actor of reference: }We identify whether the user who references the migration guide is the PR author or a reviewer. We also identify whether the actor is a human or a bot.
When a PR author references the guide, this action may indicate that the author uses the guide to understand specific migration steps required for implementation.
When a reviewer references the guide, this action may indicate that the reviewer uses the guide to verify the correctness of changes or to suggest alternatives.
When a bot references the guide, this action may indicate that the guide supports automated version updates.
By focusing on the actor, we clarify differences in usage patterns and information needs among different client types.

\noindent\textbf{(2) Location of reference: }We examine whether the link appears in the PR description or in subsequent comments.
When the link appears in the PR description, the author may use the guide to explain the migration policy.
When the link appears in comments, developers may use the guide to provide additional explanations, answer questions, or support review feedback.
This distinction allows us to determine whether developers use the guide as explanatory material for the PR or as a resource during discussion.
If we observe many comment-based references, this pattern may suggest that developers use the guide for interactive problem solving.

\noindent\textbf{(3) Level of reference: }We examine whether the referenced URL includes a fragment identifier.
A fragment identifier indicates that the user directly references a specific section.
In contrast, the absence of a fragment identifier may indicate a reference to the entire document or exploratory browsing.
This analysis allows us to evaluate how specific the required information is and how important document structure is for effective use.
We also quantify the demand for direct access to specific information within the guide.

Based on these three dimensions, we conduct an empirical analysis of the collected PRs. As supplementary information, we also record the status of each PR (merged / open / closed) and aggregate these data together with the defined dimensions.

\subsection{RQ2: Analysis of Usage Patterns}

To clarify in what context and for what purpose developers use migration guides during the development process, we conducted a manual analysis with multiple authors.

We filtered the PRs collected in RQ1 by following procedure.
First, we aim to understand how human developers use migration guides. 
Therefore, we excluded cases in which a bot referenced the migration guide.
Next, since we focus on how developers utilize migration guides for intentional use in discussion or decision making, we excluded PRs in which the link appeared only inside source code. 
Finally, we excluded PRs in which the link appeared only inside the terminal command output pasted into the PR body or comment.
We applied a translation service for PRs written in languages other than English and retained them in the set for analysis. This is because such PRs are not so many, and we aimed to preserve diversity without further reducing the sample size.

In this study, two authors independently reviewed each PR. They assigned labels based on two dimensions: the type of development and maintenance activity, and the library update status.
When disagreements occurred, a third author reviewed the conflicting cases and the corresponding labels and recommended a final decision. The first two annotators discussed the recommendation and reached full agreement.
We describe the two analyzed dimensions below.

\noindent\textbf{(4) Type of development and maintenance activity: }
We identified the purpose of development in each PR based on commit messages and discussion content.

To classify development purposes, we adopted the 14 categories used in prior work~\cite{Zampetti2017}. These categories are based on IEEE Standard for Software Maintenance~\cite{IEEE1219-1998}, the classification proposed by Swanson~\cite{Swanson1976TheDO} (Adaptive, Corrective, and Perfective), and the categories defined by Hindle et al.~\cite{Hindle2009}, including Feature Addition and Non Functional Changes. The two authors achieved 87.50\% agreement with a Cohen's kappa of 0.74, indicating substantial agreement~\cite{RichardLandis1977}.

\noindent\textbf{(5) Library update status: }
We examined whether the developer attempted to update the version of Log4j in the PR. If the developer updated the dependency, we further classified the change as a major update or a minor update.
When a developer updates a dependency, the migration guide may help resolve errors caused by breaking changes.
In contrast, a PR that does not update Log4j may represent a secondary migration triggered by updates to other libraries, or follow-up work after a previous version update. The two authors achieved 90.00\% agreement with a Cohen's kappa of 0.79, indicating substantial agreement~\cite{RichardLandis1977}.

\section{Results}
\label{sec:casestudy}

\subsection{RQ1: Quantitative Analysis of Usage Status}
We collected 64 PRs that reference the Log4j migration guide.
These PRs are based on a total of 54 repositories, covering the period from August 2015 to September 2025.
Regarding \textbf{(1) Actor of reference}, as shown in Table~\ref{tab:pr_user}, the most common case was human PR authors (38), followed by bot PR authors (15), human reviewers (10), and a bot reviewer (1).
Regarding the \textbf{(2) Location of reference}, 39 PRs cited the guide in the description, 21 cited it in comments, and 4 cited it inside source codes (Table~\ref{tab:pr_place}).
As for \textbf{(3) Level of reference} shown in Table~\ref{tab:pr_range}, 53 PRs referenced the entire migration guide, while 11 referenced a specific section.
Regarding the status of the collected PRs, 48 were merged, 8 were open, and 8 were closed.

\begin{figure}[tb]
\centering
\includegraphics[width=\columnwidth]{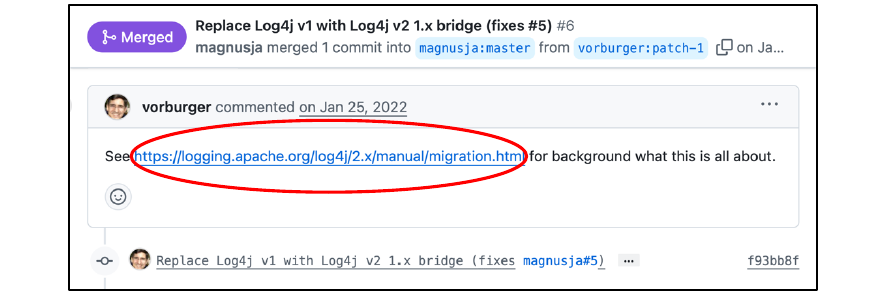}
\caption{Example PR Body with Migration Guide Reference}
\label{fig:case2}
\end{figure}

The most frequently observed pattern is a human PR author referencing the migration guide in the PR Body.
In the project shown in Figure~\ref{fig:case2},\footnote{\url{https://github.com/magnusja/java-fs/pull/6}} the author includes the URL of the migration guide, aiming to encourage the reviewer to consult the guide or to avoid repeating detailed explanations.

Among the 11 cases that referenced specific sections, frequently referenced sections include ``\texttt{\#Log4j2ConfigurationFormat},'' which supports configuration file updates, and ``\texttt{\#limitations-of-the-log4j-1-x-bridge},'' which documents tools that support migration.

Comment-based references account for 32.81\% of all cases. Among the seven cases that referenced specific sections in comments, two involved reviewers and five involved PR authors. This result suggests that developers use the migration guide for interactive problem solving during discussion.

\begin{table}[tb]
\centering
\caption{Number of Users who Referenced the Migration Guide in PRs}
\label{tab:pr_user}
\begin{tabular}{c|c|c}
\hline
 & \textbf{PR Author} & \textbf{Reviewer} \\
\hline
\textbf{Human} & 38 & 10\\
\textbf{Bot} & 15 & 1\\\hline
\end{tabular}
\end{table}
\begin{table}[tb]
\centering
\caption{Location of Migration Guide References in PRs}
\label{tab:pr_place}
\begin{tabular}{cc}
\hline
\textbf{Location} & \textbf{Number of PRs} \\
\hline
Body & 39 (60.94\%) \\
Comment & 21 (32.81\%) \\
Source Code & 4 (6.25\%) \\\hline
\textbf{Total} & \textbf{64}\\\hline
\end{tabular}
\end{table}

\begin{table}[tb]
\centering
\caption{Granularity of Migration Guide References in PRs}
\label{tab:pr_range}
\begin{tabular}{cc}
\hline
\textbf{Granularity} & \textbf{Number of PRs} \\
\hline
Entire Guide & 53 (82.81\%) \\
Specific Section & 11 (17.19\%) \\\hline
\textbf{Total} & \textbf{64}\\\hline
\end{tabular}
\end{table}

\begin{acmbox}{RQ1 Summary}
Answering \textbf{RQ1}, human PR authors most frequently reference the migration guide in the PR description (38 cases). Most references link to the entire guide (53 cases, 82.81\%), while references to specific sections remain limited (11 cases, 17.19\%).
\end{acmbox}

\subsection{RQ2: Purposes of Migration Guide Uses}

After the preliminary filtering for manual analysis, we analyzed 40 PRs.
Regarding the \textbf{(4) Type of development and maintenance activity}, the most frequent category was library compatibility, with 28 cases, which accounts for 70.00 \% of the total.
The next most frequent categories were corrective maintenance and changes outside the source code.
Regarding the \textbf{(5) Library update status}, the most frequent case was a major update, with 22 cases, which accounts for 55.00 \% of the total (Table~\ref{tab:pr_version}).
In contrast, 17 PRs did not update Log4j, which accounts for 42.50 \%.

\begin{figure}[tb]
\centering
\includegraphics[width=1.0\columnwidth]{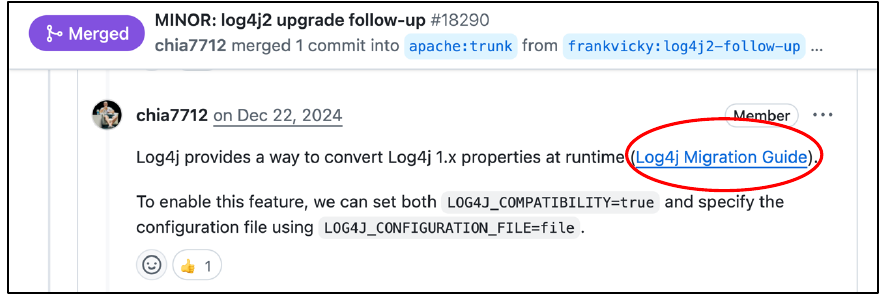}
\caption{Example Review Comment with Migration Guide Reference}
\label{fig:case1}
\end{figure}

We identified a characteristic example in the ``\texttt{apache/kafka}'' project
~\footnote{\url{https://github.com/apache/kafka/pull/18290}}.
In this project, developers performed a major Log4j update in PR \#17373. After that update, they continued to reference specific sections of the migration guide in PRs \#18290, \#18370, and \#19502, demonstrating maintenance activities after the version update (Figure~\ref{fig:case1}).


\begin{figure}[tb]
\centering
\includegraphics[width=\columnwidth, trim=10 10 10 10, clip]{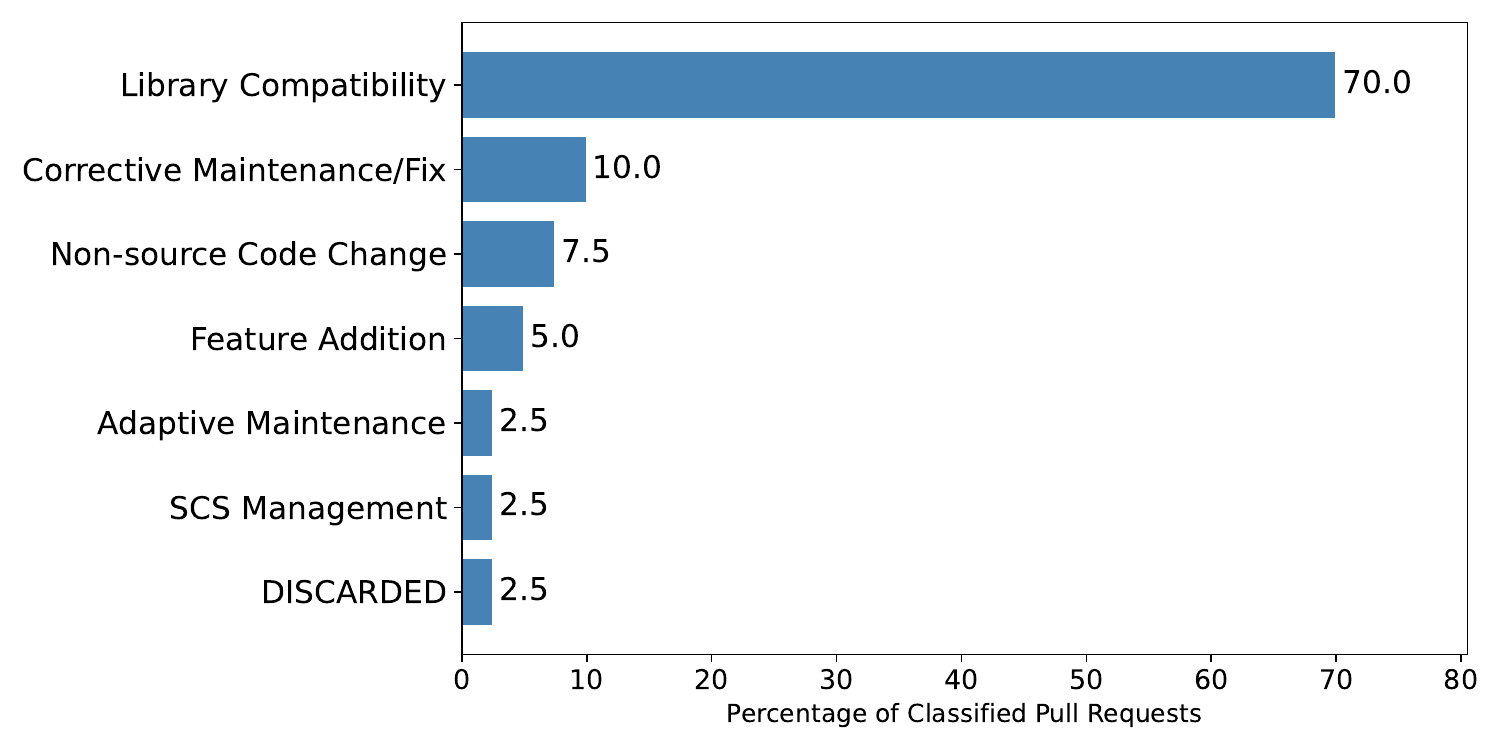}
\caption{Percentage of Classified Pull Requests by Active Category}
\label{fig:maintenance}
\end{figure}

\begin{table}[tb]
\centering
\caption{Log4j Update Status in PRs}
\label{tab:pr_version}
\begin{tabular}{cc}
\hline
\textbf{Log4j Update Status} & \textbf{Number of PRs} \\
\hline
Major Update & 22 (55.00\%) \\
Minor Update & 1 (2.50\%) \\
Non-Update & 17 (42.50\%) \\\hline
\textbf{Total} & \textbf{40}\\\hline
\end{tabular}
\end{table}

\begin{acmbox}{RQ2 Summary}
Answering \textbf{RQ2}, most PRs that reference the migration guide aim to address library compatibility (28 cases, 70.00\%). Developers reference the guide not only during major version updates (22 cases, 55.00\%) but also in PRs that do not update the version (17 cases, 42.50\%). These results suggest that the migration guide functions as a resource throughout the entire version migration process.
\end{acmbox}

\section{Discussion}
\label{sec:threats}
In this section, we discuss the implications of our findings for migration guide authors and tool developers.

\subsection{Implications for Migration Guide Authors}
Our preliminary study shows that only 27.72\% of libraries introducing breaking changes provide migration guides. This gap leaves developers without structured support and forces them to rely on release notes, which are not designed to guide migration tasks. 

Our RQ2 results reveal that 42.50\% of PRs that reference the migration guide did not involve a Log4j version update. This finding indicates that developers consult the migration guide not only during the initial migration, but also during subsequent correction and maintenance tasks. Guide authors should therefore structure their guides to clearly separate information needed at update time from information relevant to follow-up tasks. For example, dedicated sections such as ``Step-by-step migration checklist'' and ``Post-migration troubleshooting'' would allow developers to navigate the guide more efficiently depending on their current task. Our RQ1 results also show that only 17.19\% of references used fragment identifiers pointing to specific sections, suggesting that the current guide structure does not sufficiently support targeted access. Authors should design section headings and anchors with discoverability in mind to enable precise navigation.

\subsection{Implications for Tool Developers}
Our findings suggest two concrete opportunities for automated tool support. First, the low provision rate of migration guides (27.72\%) indicates a need to reduce the documentation burden on library maintainers. Tool developers should build automated generation tools that produce structured draft migration guides from API diffs, commit histories, and related sources. Recent work on LLM-based release note generation~\cite{Daneshyan2025} demonstrates the viability of this direction, and similar approaches could be applied to migration guide drafting. Second, we found that developers often reference the entire guide rather than specific sections.
This suggests that tools could help by automatically surfacing the most relevant section of a migration guide given a developer's current code change context, an IDE plugin or PR bot for example.

\section{Threats to Validity}
\label{sec:related_work}

\noindent\textbf{Construct Validity:}
We define a reference as a case in which the migration guide URL explicitly appears in the PR description or in comments.
However, this definition does not capture cases in which developers consult the guide without embedding the URL. It also does not capture cases in which developers refer to unofficial documentation.
Therefore, our approach may underestimate the actual usage of the migration guide.

\noindent\textbf{Internal Validity:}
Our analysis is observational, and we do not claim causal relationships.
Differences across PRs may be confounded by task type, repository norms, and code complexity. However, RQ1 does not control for task category.
In addition, the GitHub Search API does not search the entire index exhaustively, so some PRs may not appear in the search results.
Furthermore, we collected PRs that contain only two specific URLs, namely the former migration guide URL and its redirected page. 

\noindent\textbf{External Validity:}
Our qualitative analysis focuses only on Log4j. This focus may limit generalization to other libraries or ecosystems.
We analyzed 64 PRs that explicitly reference the migration guide. These samples may reflect project specific or community specific practices rather than broadly applicable behavior.

\section{Conclusion and Future Work}
\label{sec:conclusion}

In this study, we conducted an empirical and exploratory analysis of migration guide usage through GitHub repository mining and manual analysis.
Our results show that human PR authors most often reference the migration guide in the PR description, typically linking to the entire guide.
The main purpose of these PRs is to address library compatibility; however, developers also reference the migration guide during follow-up correction tasks after the update.
These findings suggest that migration guides serve as a useful resource throughout the continuous maintenance process, not only at the point of version migration.
We empirically demonstrate both the important role that migration guides play in the software ecosystem and the existing room for improvement in their availability and structure.

Looking ahead, two directions are particularly promising. 
First, automated tools such as LLM-based draft generators that use API diffs and commit histories could reduce the documentation burden.
Second, establishing standardized templates that clearly separate update-time and post-migration information would improve both guide quality and developer navigation.

\section*{Acknowledgement}

This research has been supported by KDDI Foundation and JSPS KAKENHI Nos. JP23K28065, JP24K14895 and JP26K21197.

\bibliographystyle{ACM-Reference-Format}
\bibliography{reference}

\end{document}